\documentclass[prb,superscriptaddress,showpacs,twocolumn]{revtex4}
\usepackage{graphicx,amsfonts}
\usepackage{epsfig,amsmath}

\usepackage{natbib,layout}

\def\figwidth{8cm}

\draft

\begin{document}

\newcommand{\atanh}
{\operatorname{atanh}}
\newcommand{\ArcTan}
{\operatorname{ArcTan}}
\newcommand{\ArcCoth}
{\operatorname{ArcCoth}}
\newcommand{\Erf}
{\operatorname{Erf}}
\newcommand{\Erfi}
{\operatorname{Erfi}}
\newcommand{\Ei}
{\operatorname{Ei}}

\title{Non equilibrium dynamics below the super-roughening transition.} 
\author{Gregory Schehr}
\affiliation{Theoretische Physik Universit\"at des Saarlandes
66041 Saarbr\"ucken Germany}
\author{Heiko Rieger}
\affiliation{Theoretische Physik Universit\"at des Saarlandes
66041 Saarbr\"ucken Germany}

\draft

\date{\today}
\begin{abstract}
The non equilibrium relaxational dynamics of the solid on solid model
on a disordered substrate and the Sine Gordon model with random phase
shifts is studied numerically. Close to the super-roughening temperature
$T_g$ our results for the autocorrelations, spatial correlations and
response function as well as for the fluctuation dissipation ratio
(FDR) agree well with the prediction of a recent one loop RG
calculation, whereas deep in the glassy low temperature phase
substantial deviations occur. The change in the low temperature
behavior of these quantities compared with the RG predictions is shown
to be contained in a change of the functional temperature dependence
of the dynamical exponent $z(T)$, which relates the age $t$ of the
system with a length scale ${\cal L}(t)$: $z(T)$ changes from a linear
$T$-dependence close to $T_g$ to a $1/T$-behavior far away from
$T_g$. By identifying spatial domains as connected patches of the
exactly computable ground states of the system we demonstrate that the
growing length scale ${\cal L}(t)$ is the characteristic size of
thermally fluctuating clusters around ``typical'' long-lived
configurations.
\end{abstract}
\maketitle


\section{INTRODUCTION.}

Despite many efforts the understanding of non-equilibrium dynamics of
disordered and glassy systems in finite dimensions remains a
challenging problem. In particular in glasses and spin glasses the
aging process displays a very rich phenomenology demanding new
theoretical concepts \cite{leto_leshouches}. But already less complex
--- and apparently less glassy --- systems, like disordered but
non-frustrated systems \cite{random-ising} or even pure systems 
\cite{cugliandolo_pure} reveal interesting and
unexpected aging phenomena. One of the most intruiging questions in
this context is whether the out-of-equilibrium dynamics is essentially
fully determined by a coarsening process (a question that even arises
in the more complex spin glass situation \cite{nao-rie-rev}),
describable by a growing length scale that characterizes essentially
all out-of-equilibrium processes. In this paper we will consider a
disordered system in which this question has not been clarified yet ---
and for 
which the answer that we find will reveal a non-standard scenario.

Among glassy systems, there is a wide interest in disordered elastic
systems, which cover a wide range of physical situations ranging from
vortex lattices in superconductors \cite{revuevortex}, interfaces in
disordered magnets \cite{ferre,krusin} or electron glasses
\cite{coulombglass} for which nonequilibrium effects are
experimentally relevant.  Here, we investigate the non
equilibrium relaxational dynamics of a solid-on-solid (SOS) model on a
disordered substrate, defined on a two dimensional square lattice and
described by the following elastic Hamiltonian in terms of height
variables $h_i$
\begin{eqnarray}
H_{\text{SOS}} = \sum_{\langle ij \rangle} (h_i - h_j)^2 \quad , \quad
 h_i \equiv n_i + d_i
 \label{Def_SOS} 
\end{eqnarray}
where $n_i$ are unbounded discrete variables, {\it i.e.} $n_i \in \{0,
\pm 1, \pm 2,..\}$ and $d_i \in [0,1[$ are uniformly distributed quenched
random offsets, uncorrelated from site to site. 
In the absence of disorder, {\it i.e.} $d_i=0$, the model exhibits a
roughening transition in the 
same universality class as the Kosterlitz-Thouless transition
\cite{nozieres_roughening},    
at a temperature $T_r$ separating a flat phase at low $T$ from a
logarithmically (thermally) rough one above $T_r$. The presence of
disorder is known to modify significantly the nature of the
transition \cite{toner,tsai_super_rough,fisherhwa}. The so-called
{\it superroughening} transition occurs at a temperature $T_g =
T_r/2 = 2/\pi$. Above $T_g$, where the disorder is irrelevant on
large length scales, the surface is logarithmically rough again, although
below $T_g$ the system exhibits a glassy phase where the pinning
disorder induces a stronger roughness of the interface. 
In the continuum
limit, near $T_g$, this SOS model (\ref{Def_SOS}) is in the same
(equilibrium) universality class as the Sine-Gordon model with 
random phase shifts, the so-called Cardy-Ostlund (CO) model \cite{co}
\begin{eqnarray}
H_{\text{{CO}}} =  \int d^2x (\nabla
\varphi(x))^2 - \Delta \cos{ (2\pi (\varphi(x) - \xi(x)))} \label{Def_SG}
\end{eqnarray}
where $\varphi(x) \in ]-\infty,+\infty[$ is a continuous variable 
and  $\xi(x) \in [0,2\pi[$ is a
uniformly distributed quenched random phase variable, uncorrelated
from site to site, $\Delta$ being the strength of the disorder. The
model (\ref{Def_SG}) arises in various contexts like the XY model in
a random magnetic field (without vortices) or in vortex physics where
it describes a $2$-d array of flux lines pinned by point like disorder
\cite{natter1}. The low temperature glassy phase ({\it i.e.} below
$T_g$) of these models
(\ref{Def_SOS}, \ref{Def_SG}) is described by a finite temperature
fixed point associated with a free energy exponent $\theta=0$, which is 
an exact statement due to the statistical tilt symmetry
\cite{schulz_STS}. 

Although these models have been extensively studied, both analytically
\cite{carpentier_co} and numerically
\cite{simu_tc, rieger_gs_sos, middleton_gs_sos}, these works have mainly 
focused on the equilibrium properties. Among them the static roughness
of the interface has been investigated thoroughly and for the dynamics
the dynamical exponent $z$ \cite{tsai_super_rough,gold}.  The latter
was found to depend continuously on $T$ and computed using the
renormalization group (RG) up to one loop in the vicinity of $T_g$,
where the fixed point is controlled by the small parameter $\tau =
(T_g-T)/T_g$.  Only recently, the non-equilibrium relaxational
dynamics (defined by a Langevin equation) of the Cardy-Ostlund model
(\ref{Def_SG}) was investigated analytically \cite{schehr_co_prl} in
the perturbative regime ($\tau\ll1$). Using the RG this study focused
on the the two-times ($t,t_w$) correlation and response functions. The
autocorrelation and local response function were found to scale as
$t/t_w$ and characterized by asymptotically algebraic scaling
functions with an associated decay exponent that depends continuously
on $T$ and was calculated perturbatively up to one loop
order. Finally, the associated fluctuation dissipation ratio (FDR) in
the large time separation limit was found to be non-trivial and also
$T$-dependent.

In this paper we intend first to test numerically this analysis near
$T_g$, then to go beyond the perturbative regime and explore the low
$T$ dynamics where one expects to observe a stronger signature of the
logarithmic free energy landscape
\cite{marginal_freezing} as suggested by the static value of $\theta = 0$. 
Furthermore, having determined these different non equilibrium dynamical
properties, we propose to relate them to a real space analysis 
of the equilibration process of the thermal fluctuations in the system. 
Their quantitatively precise study is possible due to an algorithm 
\cite{rieger_gs_sos, middleton_gs_sos} that allows one to compute the 
exact ground state of (\ref{Def_SOS}). 
 
The outline of the paper is as follows. In Section II, we give some
details of our simulations and present the definitions of the dynamical
two times quantities we will focus on. In Section III, we present our
numerical results for these quantities, and establish a comparison with
the analytical predictions of \cite{schehr_co_prl} (some details of this
comparison are left in Appendix A). Section IV is devoted
to a physical discussion, based an aging scenario in real space. Finally
we draw our conclusions in the last section.

\section{SIMULATIONS and DEFINITIONS.}

We perform a numerical study of the nonequilibrium relaxational
dynamics of these models ({\ref{Def_SOS}, \ref{Def_SG}}) on a $2$-d
square lattice with periodic boundary conditions using a standard
Monte Carlo algorithm. Although the SOS model is by definition a
discrete model, the CO model (\ref{Def_SG}), which is a continuous
one, needs to be discretized for the purpose of the simulation. We
will use the discretized version of the gradient in (\ref{Def_SG}),
with $\varphi(x) \to \varphi_i$, $i$ being the site index. The value
of the displacement field $\varphi_i$ is itself discretized into
$4096$ intervals of width $\Delta \varphi$ between $\pm 4$.  Except
when we explicitly mention it, the system is initially prepared in a
flat initial condition ($n_i(t=0) = 0$ or $\varphi_i(t=0) = 0$).  At
each time step, one site is randomly chosen and a move $n_i \to n_i +
1$ or $n_i \to n_i - 1$ is proposed with equal probability (for the CO
model, the field $\varphi_i$ is incremented or decremented by an
amount $\Delta \varphi$). This move is then accepted or rejected
according to the heat bath rule.  Our data were obtained for a lattice
of linear size $L = 
64$ or $L=128$, and a time unit corresponds to $L^2$ time steps.

We will first study the connected autocorrelation
function ${\cal C}(t,t_w)$ 
\begin{eqnarray}
{\cal C}(t,t_w) = \frac{1}{L^2} \sum_i \overline{\langle h_i(t)
  h_i(t_w) \rangle -  
\langle h_i(t) \rangle \langle h_i(t_w) \rangle}\;, \label{Def_Autocorrel}
\end{eqnarray}
which is a two-time quantity allowing to characterize aging
properties. Then we will consider the spatial (2-point) connected
correlation function 
\begin{eqnarray}
C(r,t) = \frac{1}{L^2} \sum_i \overline{\langle h_i(t) h_{i+r}(t) \rangle - 
\langle h_i(t) \rangle \langle h_{i+r}(t) \rangle}
\label{Def_Struct_Fact} 
\end{eqnarray}
from which we measure the dynamical exponent $z$. In
(\ref{Def_Autocorrel}, \ref{Def_Struct_Fact}), $\langle...\rangle$ and
$\overline{...}$ mean an average over the thermal noise and
respectively over the disorder.  When studying the CO model
(\ref{Def_SG}) the corresponding correlation functions are defined by
Eq. (\ref{Def_Autocorrel}, \ref{Def_Struct_Fact}) with the
substitution $h_i(t) \to \varphi_i(t)$.

These two quantities (\ref{Def_Autocorrel}, \ref{Def_Struct_Fact}) are
straightforwardly computed from our simulation which stores at each
time step $t$ the value of the height field $h_i(t)$ on each site
$i$. Typically, in our simulations we compute ${\cal C}(t,t_w)$ by
averaging over 64 (32) different realizations of the thermal noise for
a given configuration of the disorder and then averaging over 256
(128) different disorder samples for $L=64$ (respectively $L=128$). We
observed that the main fluctuations in the computation of the
correlations were coming from the average over the
disorder. Therefore, we have estimated the error-bars from the sample
to sample fluctuations of the thermal average value in
(\ref{Def_Autocorrel},\ref{Def_Struct_Fact}).

We are also interested in the violation of the
fluctuation dissipation theorem (FDT) associated with {\it local}
fluctuations (\ref{Def_Autocorrel}) for which we have to consider  
the associated local linear response ${\cal R}(t,t_w)$ 
\begin{eqnarray}
{\cal R}(t,t_w) = \overline{\left\langle \frac{\delta h_i(t)}{\delta f_i(t_w)}
\right \rangle}\;,
\end{eqnarray}
where $f_i(t_w)$ being an infinitesimal force applied at site $i$ at
time $t_w$. The dynamical rules are then modified by adding a term
$-\sum_i f_i n_i $ to the original Hamiltonian
Eq. (\ref{Def_SOS}). Numerically it is more convenient to calculate
instead the integrated response
\begin{eqnarray}
\rho(t,t_w) = \int_0^{t_w} ds {\cal R}(t,s) \label{Def_rho}\;.
\end{eqnarray}
In order to isolate the diagonal component of the response function, we
used the standard strategy
\cite{barrat_response,godreche_response}: we simulate two 
replicas of the system, one without an applied force and another in
which we apply a spatially random force to the system from time $t=0$
to time $t=t_w$. This force field is of the form $f_i = f_0
\epsilon_i$, with a constant small amplitude $f_0$ and a quenched
random modulation $\epsilon_i = \pm 1$ with equal probability,
independently at each site $i$. The integrated response $\rho(t,t_w)$
is then computed as
\begin{eqnarray}
\rho(t,t_w) = \frac{1}{L^2} \sum_i \overline{\frac{\langle h_i(t)
 \rangle_{f_i} - 
 \langle h_i(t) 
 \rangle}{f_i}}  \quad, \quad t > t_w\;, \label{Def_Rho}
\end{eqnarray}
where $\langle h_i(t) \rangle_{f_i}$ means the thermal average in the
presence of the force field $f_i$.  We have used a numerical value of
$f_0 = 0.3$ and have checked that we were indeed probing the linear
response regime. Our numerical data for $\rho(t,t_w)$ are averaged
over $64$ (32) independent thermal realizations for a given disorder
configuration and the random fields $f_i$ for $L=64$ (resp. $L=128$)
and then averaged over $512$ different disorder realizations.  The
error-bars are estimated in the same way as for the correlation
functions.  We point out that instead of $\rho(t,t_w)$ many studies,
{\it e.g} in spin glasses, focus on $\chi(t,t_w) =
\int_{t_w}^t ds {\cal R}(t,s)$. In our model in which one time
quantities such as ${\cal C}(t,t)$ grow without bounds when $t$
increases there may be a regime in which the integral over $s$ in the
definition of $\chi(t,t_w)$ is actually dominated by the latest times
$s$ \cite{response_one_time} and thus depends only very weakly on the
waiting time $t_w$. Therefore, in order to disentangle the off-diagonal
part of the response itself the computation of ${\rho}(t,t_w)$
(\ref{Def_Rho}), which does not suffer from the aforementioned peculiarity,
is better suited.

When the system is in equilibrium the dynamics is time
translation invariant (TTI) and two-times quantities like ${\cal
 C}(t,t_w)$ or $\rho(t,t_w)$ depend only on the time 
difference $t-t_w$. Moreover, ${\cal C}(t,t_w)$ and the response
${\cal R}(t,t_w)$ are related by the Fluctuation Dissipation Ratio (FDT):
\begin{eqnarray}\label{FDT}
\partial_{t_w} {\cal C}(t,t_w) = T {\cal R}(t,t_w)
\end{eqnarray}
When the system is not in equilibrium, these properties do not hold any
more and it has been proposed to generalize the FDT to nonequilibrium 
situations by defining a Fluctuation Dissipation Ratio (FDR)
${X}(t,t_w)$\cite{leto_leshouches,cugliandolo_fdr}:  
\begin{eqnarray}
\frac{T}{X(t,t_w)} = \frac{\partial_{t_w}{\cal C}(t,t_w)}{{\cal
    R}(t,t_w)} \label{Def_FDR}
\end{eqnarray}
such that $X(t,t_w) = 1$ in equilibrium (\ref{FDT}) and any deviation
from unity 
being a signature of an out of equilibrium situation. In this paper, we
will investigate this FDR (\ref{Def_FDR}) for the (nonequilibrium)
relaxational dynamics following a sudden quench at $t=0$. 
Of particular interest is the limiting value 
\begin{eqnarray}
X^{\infty} = \lim_{t_w \to \infty}
\lim_{t \to \infty} X(t,t_w)\;,\label{xinf} 
\end{eqnarray}

\section{RESULTS}

\subsection{Correlation function.}

\subsubsection{Autocorrelation function.}
\begin{figure}
\includegraphics[width=\figwidth]{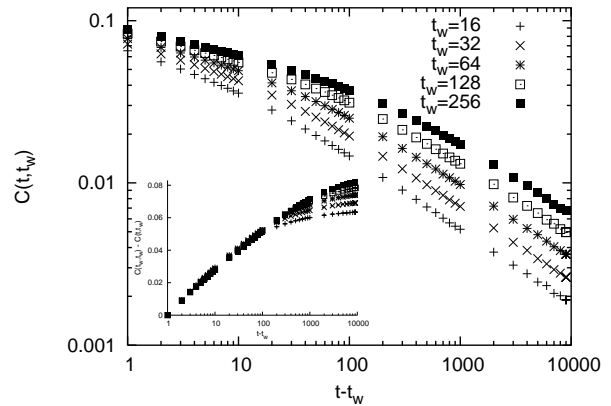} 
\caption{Connected correlation function ${\cal C}(t,t_w)$ as a function
 of $t-t_w$ for different waiting times $t_w$. The inset shows the plot
 of $C(t_w,t_w) - C(t,t_w)$ as a function of $t-t_w$, for the same
 different waiting times, which exhibits the ``quasi-equilibrium''
 regime. Here, $T=0.63
 \,T_g$.}\label{AutoCorrel0.4} 
\end{figure}

  %
Figure \ref{AutoCorrel0.4} shows the decay of the connected correlation
function ${\cal C}(t,t_w)$ for different waiting times $t_w$ and for a
temperature $T = 0.63 \,T_g$ : they show a clear $t_w$ dependence.
We notice that the quantity ${\cal C}(t_w,t_w)$ depends also on $t_w$,
before saturating to its equilibrium value for $t_w \to \infty$ (which
depends on the system size $L$). This
explains why one does not observe a ``quasi-equilibrium'' regime,
where ${\cal C}(t,t_w) \equiv {\cal C}(t-t_w)$ when $t-t_w \ll t_w$ for
the relatively small waiting times showed in Fig. \ref{AutoCorrel0.4}. This
``quasi-equilibrium'' regime can however be observed if we plot 
${\cal C}(t_w,t_w) - {\cal C}(t,t_w)$, as shown on the inset of Fig. 1. 

\begin{figure}
\centering
\includegraphics[width=\figwidth]{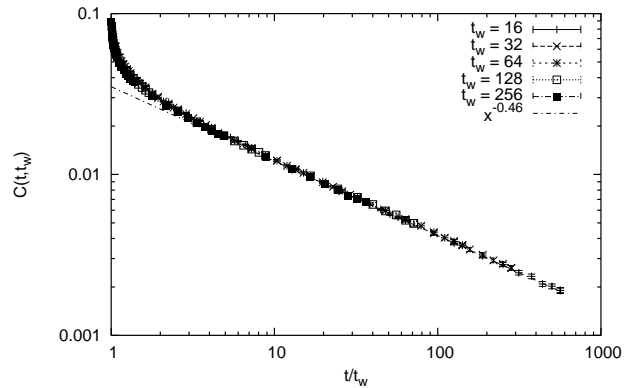}
\caption{Connected correlation function ${\cal C}(t,t_w)$ as a function
 of $t/t_w$ for different waiting times at temperature $T=0.63\,T_g$.
 The dotted line is the result of the fit (\ref{AutoCorrelAsympt})
 taking into account the data points with $t/t_w >
 10$.}\label{RescAutoCorrel0.4} 
\end{figure}

\begin{figure}
\centering
\includegraphics[width=\figwidth]{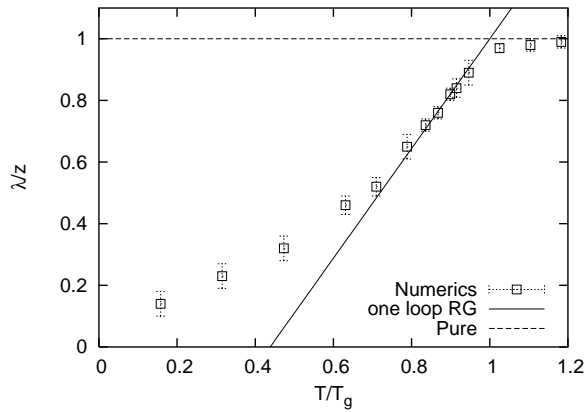}
\caption{Decay exponent $\lambda/z$ as a function of
 $T/T_g$. The dashed lined indicates the exact value for $T>T_g$. The
 solid line shows the result of the one loop RG \cite{schehr_co_prl},
 given in (\ref{one_loop_lambda}). Importantly, this curve is drawn
 without any fitting parameter, $T_g = 2/\pi$ being exactly
 known.}\label{Decay_Exp}   
\end{figure}

In the aging regime, for $t-t_w \sim {\cal O}(t_w)$, these curves  
for different waiting times $t_w$ fall on a single master curve when
we plot ${\cal C}(t,t_w)$ as a function of $t/t_w$
(Fig. \ref{RescAutoCorrel0.4}). 
In the large time separation regime $t \gg t_w$ these data are well
fitted by a power law decay 
\begin{eqnarray}
{\cal C}(t,t_w) \sim \left( \frac{t}{t_w}\right)^{-\lambda/z} \quad,
 \quad t \gg t_w \label{AutoCorrelAsympt}
\end{eqnarray}
Notice however that one can not exclude logarithmic corrections at low
temperature where the decay exponent becomes very small. In
Fig. \ref{Decay_Exp}, we plot the value of the decay exponent
$\lambda/z$ for different temperatures.  In the high temperature
phase, $T>T_g$, where $\lambda=z=2$ one expects $\lambda/z = 1$ {\it
independent} of $T$ (notice that the high temperature phase is 
critical and as such also displays aging behavior
\cite{cugliandolo_pure, berthier_xy}). For $T<T_g$ the presence of
disorder reduces the decay exponent $\lambda/z$ which now depends
continuously on temperature. In the vicinity of $T_g$ one observes a
rather good agreement with the perturbative RG computation to one loop
\cite{schehr_co_prl}:
\begin{eqnarray}\label{one_loop_lambda}
\frac{\lambda}{z} = 1 - e^{\gamma_E} \tau + {\cal O}(\tau^2)
\end{eqnarray}
where $\gamma_E = 0.577216$ is the Euler constant. With the 
RG result $z = 2 + 2 e^{\gamma_E} \tau + {\cal O}(\tau^2)$ 
this corresponds to $\lambda = 2 + {\cal O}(\tau^2)$.  

\begin{figure}
\centering
\includegraphics[width=\figwidth]{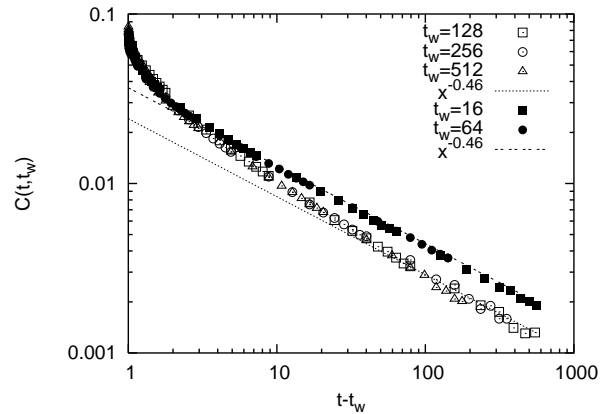}    
\caption{Connected correlation function ${\cal C}(t,t_w)$ obtained with
 the SOS model (filled symbols) and with the CO hamiltonian (open
 symbols) as a function of $t/t_w$ for 
 different $t_w$. Here $T=0.63\,T_g$.
}\label{Comp_SOS_SG} 
\end{figure}

Notice that the simulations near $T_g$, $T/T_g \gtrsim 0.8$, {\it i.e.}
in the weak 
disorder regime, have been performed using
the random phase Sine-Gordon formulation (\ref{Def_SG}) of the SOS
model, for which the asymptotic regime is faster reached for these
temperatures. The inverse is of course true at low temperature. 
When it was possible, we have compared for a given
temperature the asymptotic
properties of ${\cal C}(t,t_w)$ using the SOS model (\ref{Def_SOS}) and
the CO model (\ref{Def_SG}). We show the result of this comparison for
$T=0.63\,T_g$ in Fig. \ref{Comp_SOS_SG}.

One observes that both formulations are in good agreement concerning
the $t/t_w$ scaling form and are in a reasonable agreement concerning
the value of the exponent $\lambda/z$, thus confirming the
universality of this property. However, the amplitude itself does not
seem to be universal.

\begin{figure}
\includegraphics[width=\figwidth]{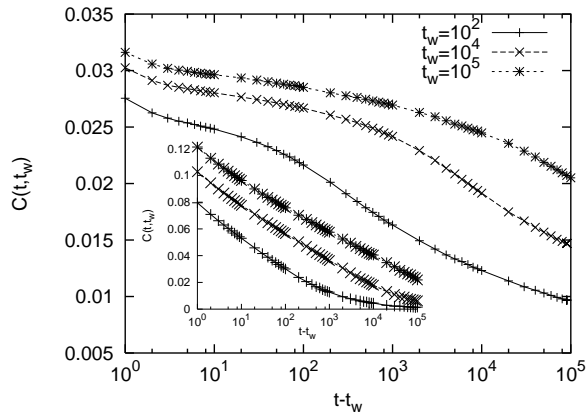}  
\caption{Autocorrelation function ${\cal C}(t,t_w)$ as a function of
 $t-t_w$ for different large waiting times $t_w$, at (very) low
 temperature, $T=0.15 \,T_g$. For short $t-t_w$, this quantity shows an
 inflection point.  The inset shows the same quantity for $T=0.63 \,T_g$,
 which exhibits a qualitatively different behavior for $t-t_w \ll
 t_w$. 
 }\label{Correl_low_T}
\end{figure}

At lower temperature the perturbative calculation fails to predict the
correct behavior of $\lambda/z$: in Fig. \ref{Decay_Exp} we observe a
change in its $T$-dependence below $T\approx0.8\,T_g$.  In this regime
one obtains a good fit of the decay exponent by
\begin{eqnarray}\label{fit_lambda}
\frac{\lambda}{z} \sim A_{\lambda/z} T \quad, \quad A_{\lambda/z} =
0.85 \pm 0.04  
\end{eqnarray}
If one naively assumes that the one loop RG calculation $\lambda = 2$
is still valid at low temperature, this would already indicate a $1/T$
behavior of the dynamical exponent $z$. We will come later to this
point where we explicitly compute this exponent $z$. Indeed, this
scaling form (\ref{AutoCorrelAsympt}) can be written as
\begin{eqnarray}\label{AutoCorrel_Coars}
{\cal C}(t,t_w) \sim \left(\frac{{\cal L}(t)}{{\cal L}(t_w)}
\right)^{-\lambda} \quad, \quad {\cal L}(t) \sim t^{1/z}
\end{eqnarray}
thus defining a length scale ${\cal L}(t)$ which can be further
analyzed by measuring how the spatial correlations are growing in the
system (see the next paragraph). The functional shape of ${\cal
C}(t,t_w)$ that we determined suggests that its $T$-dependence is
mainly contained in the decay exponent within the the aging regime
where $(t-t_w)\sim {\cal O}(t_w)$.  It is remarkable that its most
prominent feature, the $t/t_w$ scaling and the asymptotically
algebraic scaling form with a $T$-dependent decay exponent, is already
captured by the one loop RG calculation of \cite{schehr_co_prl}. By
contrast, one observes that the ``quasi-equilibrium'' regime
$(t-t_w) \ll t_w$ shows a much stronger $T$-dependence. At low
temperature $T \lesssim T_g/2 $ the autocorrelation function ${\cal
C}(t,t_w)$ displays an inflection point at small time difference
$t-t_w$.  In Fig. \ref{Correl_low_T}, where ${\cal C}(t,t_w)$ as a
function of $t-t_w$ is shown in a linear-log plot for different large
waiting times $t_w$, one observes a qualitative change of behavior which
could suggest a finite limiting value $\lim_{t \to \infty} \lim_{t_w \to
\infty} {\cal C}(t,t_w)$. However, on the time scales explored here, we
have not identified a clear signature of such a behavior. Nevertheless, 
this point deserves further investigations of the equilibrium
properties at low temperature, where some discrepancies between
numerics \cite{rieger_gs_sos,middleton_gs_sos} and 
analytical predictions \cite{ludwig} were already found.

\subsubsection{$2$-point correlation function.}

\begin{figure}
\includegraphics[width=\figwidth]{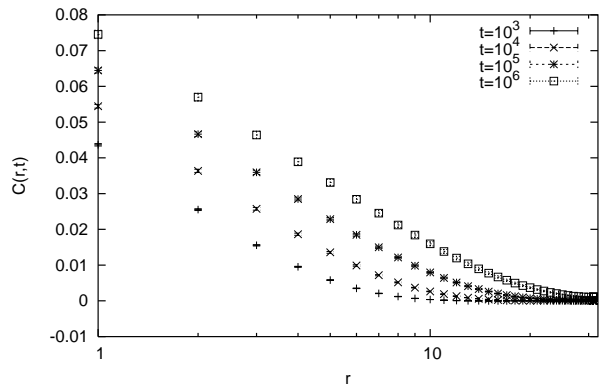}
\caption{Spatial connected correlation function $C(r,t)$ as a function
 of $r$ for different times $t$. Here $T=0.47 \,T_g$.}\label{StructFact0.3}
\end{figure}

In Fig. \ref{StructFact0.3} we show the $2-$point
correlation function (\ref{Def_Struct_Fact}) for a temperature 
$T=0.47\,T_g$ (and $L=64$) for different times $t$. 

\begin{figure}
\includegraphics[width=\figwidth]{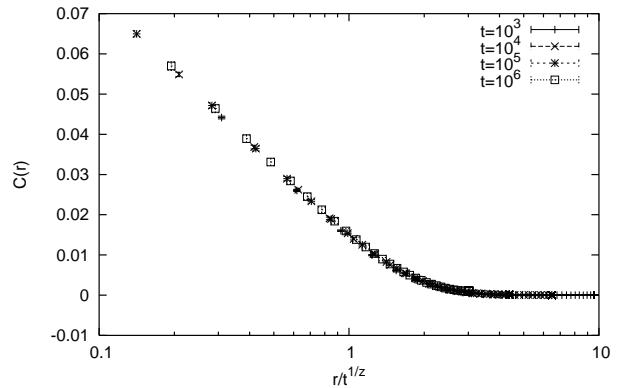}
\caption{Spatial connected correlation function $C(r,t)$ as a function
 of $r/t^{1/z}$ with $1/z=0.17 \pm 0.01$ for different times $t$. Here
 $T=0.47 \,T_g$}\label{RescStructFact0.3} 
\end{figure}

As $t$ grows spatial correlations develop in the system. More
precisely, as shown in Fig. \ref{RescStructFact0.3}, $C(r,t)$ 
scales as
\begin{eqnarray}
C(r,t) = {\cal F}\left( \frac{r}{{\cal L}(t)}  \right) \quad, \quad
 {\cal L}(t) \sim t^{1/z}\;. \label{Scaling_Struct_Fact}
\end{eqnarray}
The value of $z$ that gives the best data collapse leads to our first
estimate of the dynamical exponent. The logarithmic behavior for 
$r \ll {\cal L}(t)$, $C(r,t) \sim \ln{{\cal L}(t)/r}$ is in agreement
with the constraint imposed by the statistical tilt symmetry (STS)
\cite{schulz_STS} which
fixes the equilibrium behavior of the connected $2$-point correlation
function to
\begin{eqnarray}
\lim_{t \to \infty} C(r,t) \sim -\frac{2}{(2 \pi)^2} \frac{T}{T_g}
\ln{r}\;, \label{sos} 
\end{eqnarray}  
which is identical with the pure (i.e. disorder-free behavior).
We also checked that the amplitude of the logarithmic behavior of
$C(r,t)$ for $r/{\cal L}(t) \ll 1$ is in good agreement (within 
a few percents) with Eq. (\ref{sos}).

\subsubsection{Dynamical exponent.}

Another way to estimate the dynamical exponent is to determine the
time dependent  
length scale ${\cal L}(t)$ itself. For that purpose, and given the
scaling form 
previously computed (\ref{Scaling_Struct_Fact}), we estimate ${\cal
  L}(t)$ via a 
the space integral of the spatial correlations \cite{kisker_noneq_sg}:
\begin{equation}
\int_0^{L/2} \!\!dr\, C(r,t) = \int_0^{L/2}\!\! dr\, {\cal F}(r/{\cal L}(t)) 
\sim {\cal L}(t) \int_0^{\infty} du\, {\cal F}(u) \label{Calc_L}\,,
\end{equation}
where we assumed in the last step that $L/{\cal L}(t) \ll 1$ (which is
indeed the case on the time scales considered here) and that $C(r,t)$
decays sufficiently fast at large $r$ (we checked that it actually
decays exponentially). Notice also that the sum in (\ref{Calc_L}) is
bounded to $L/2$ due to periodic boundary conditions.
\begin{figure}
\includegraphics[width=\figwidth]{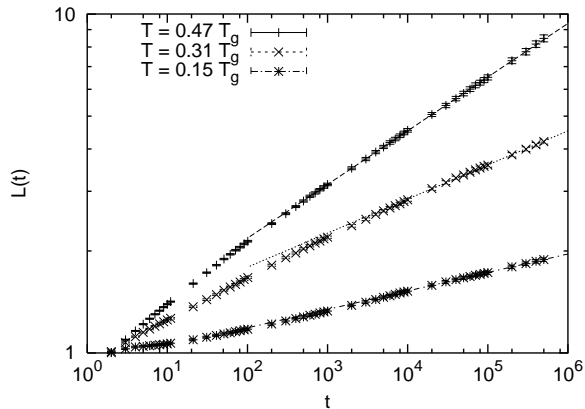}
\caption{Growing length scale ${\cal L}(t)$ computed from (\ref{Calc_L})
 for different temperatures. The solid lines are guides to the eyes. 
 }\label{Length}  
\end{figure}
In Fig. \ref{Length} we showed the value of ${\cal L}(t)$ computed
with (\ref{Calc_L}) for different temperatures. One obtains a rather
good fit of these curves (Fig. \ref{Length}) by a power law ${\cal
L}(t)\sim t^{1/z(T)}$, thus obtaining a value of the $T$ dependent
dynamical exponent in good agreement with the value obtained by
collapsing the different curves in Fig. \ref{RescStructFact0.3}. One
notices also that ${\cal L}(t)$ approaches an algebraic growth after a
pre-asymptotic regime which increases with decreasing temperature.
Fig. \ref{Dyn_Exp} shows our estimate for $1/z(T)$ as a function of
$T$.
\begin{figure}
\includegraphics[width=\figwidth]{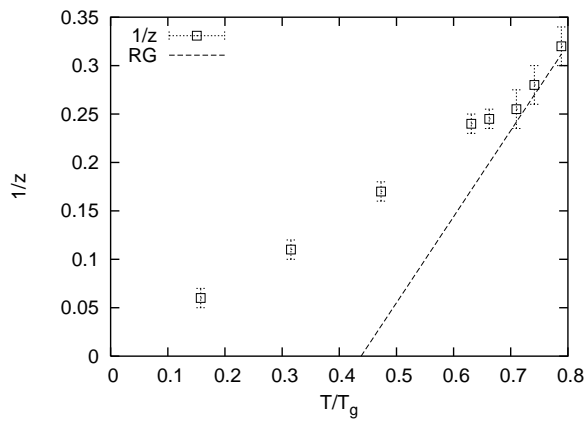}
\caption{$1/z(T)$ as a function of $T/T_g$. The dashed line which shows
 the result of the one loop RG \cite{tsai_super_rough,gold} is drawn
 without any fitting parameter.}\label{Dyn_Exp} 
\end{figure}
As expected, the dynamical exponent is a decreasing function of the
temperature. One expects that $z=2$ for $T>T_g$ and that it becomes
$T$-dependent below $T_g$ with $z = 2 + 2e^{\gamma_E}\tau + {\cal
O}(\tau^2)$ as predicted by a one loop RG calculation
\cite{tsai_super_rough,gold}.  At high temperature $T>T_g$ and in the
vicinity of $T_g^-$, it is numerically rather difficult to extract a
reliable estimate for the dynamical exponent from
(\ref{Scaling_Struct_Fact}) or (\ref{Calc_L}) due to finite size
effects. Therefore we restrict ourselves here to lower
temperatures $T < 0.8 \,T_g$. For temperature $T\gtrsim 0.7 \,T_g$, the
value of $z$ is still in reasonable agreement with the RG
prediction. Around the value $T^*
\simeq 0.63 \,T_g$, where $z \simeq 4$, the curve $1/z(T)$
shows an inflection point, below which $1/z$ decreases linearly with
$T$. In this regime, $z(T)$ is well fitted by
\begin{eqnarray}
z(T) \sim 4 \frac{T^*}{T} 
\quad{\rm for}T\le T^*
\label{z_1oT}
\end{eqnarray}     
which, given (\ref{fit_lambda}), shows also that $\lambda \simeq 2$ is
still a good estimate 
a low $T$. This behavior $z \propto 1/T$ is compatible with an activated
dynamics 
over logarithmic barriers, {\it i.e.} an Arrhenius type behavior
$t_{\text{typ}} \sim e^{{B_L}_{\text{typ}}/T}$ with
$B_{L_{\text{typ}}} \sim \log{L_{\text{typ}}}$. Assuming that the
largest barriers, which dominate the low temperature dynamics,
encountered in this non equilibrium relaxation process have the same
scaling that the equilibrium ones, this logarithmic behavior is also
consistent with a free energy exponent $\theta = 0$
\cite{drossel_theta_psi}. Interestingly, this change of behavior of
$z$ at a value of $z_c = 4$ above which $z\propto 1/T$ (\ref{z_1oT})
is reminiscent of the related case of a particle in a one dimensional
disordered potential with logarithmic correlations where such a
behavior was obtained analytically \cite{marginal_freezing}. It should
be mentioned that a dynamical exponent that varies like $1/T$ has also
be found in other disordered systems like in spin glasses
\cite{kisker_noneq_sg,parisi-prl} and in random ferromagnets
\cite{paul-rbfm}. Finally, although (\ref{z_1oT}) suggests the
existence of a well defined {\it typical} relaxation time, one expects
the full distribution of the barrier heights to be very broad
\cite{balents_barriers}, and needs probably further work to be
investigated.

\subsection{Integrated response function.}

\begin{figure}
\includegraphics[width=\figwidth]
 {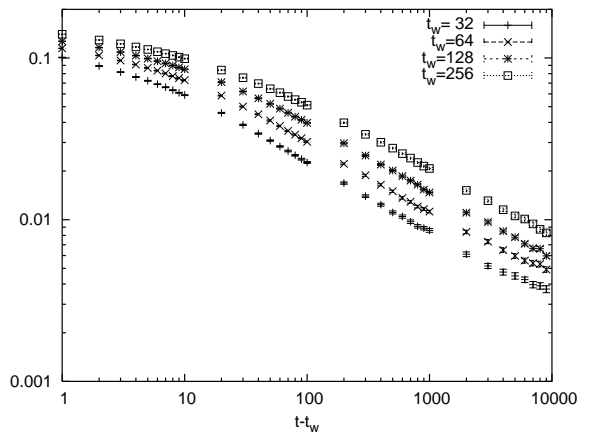}    
\caption{Integrated response function $\rho(t,t_w)$ as a function of
 $t-t_w$ for different waiting times $t_w$. Here
 $T=0.47\,T_g$.}\label{Resp0.3}
\end{figure}

In this section, we focus on the integrated response 
(\ref{Def_Rho}). In Fig. (\ref{Resp0.3}) we show a plot of $\rho(t,t_w)$
as a function of the time difference $t-t_w$ for different waiting times
$t_w$. Here too, one observes a clear waiting time dependence.  

\begin{figure}
\includegraphics[width=\figwidth]
 {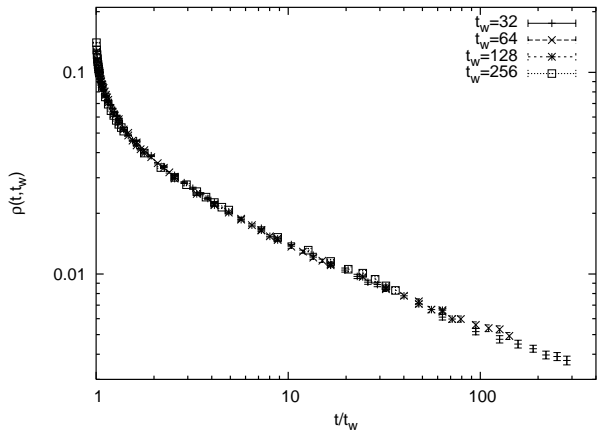}    
\caption{Integrated response function $\rho(t,t_w)$ as a function of
 $t/t_w$ for different waiting times $t_w$ at $T=0.47
 \,T_g$.}\label{RescResp0.3}
\end{figure}

These curves for different waiting times $t_w$ fall on a single master curve 
if one plots them as a function of $t/t_w$, as shown in
Fig. \ref{RescResp0.3}.
As suggested on this log-log plot (Fig. \ref{RescResp0.3}),
$\rho(t,t_w)$ takes the following power law decay
\begin{eqnarray}
\rho(t,t_w) \sim \left( \frac{t}{t_w}\right)^{-\lambda/z} \quad, \quad t
 \gg t_w
\end{eqnarray}
Notice that the decay exponent, within the accuracy of the data
presented here, is the same as the one of the corresponding
autocorrelation function ${\cal C}(t,t_w)$ (\ref{AutoCorrelAsympt}). This
$t/t_w$ scaling form, together with the relation between the decay
exponent of $\rho(t,t_w)$ and ${\cal C}(t,t_w)$ are also fully
compatible with previous one loop RG calculations. As we will see, this
has important implications for the FDR as discussed in the next paragraph.

\subsection{Fluctuation dissipation ratio.}

\begin{figure}
\includegraphics[width=\figwidth]
 {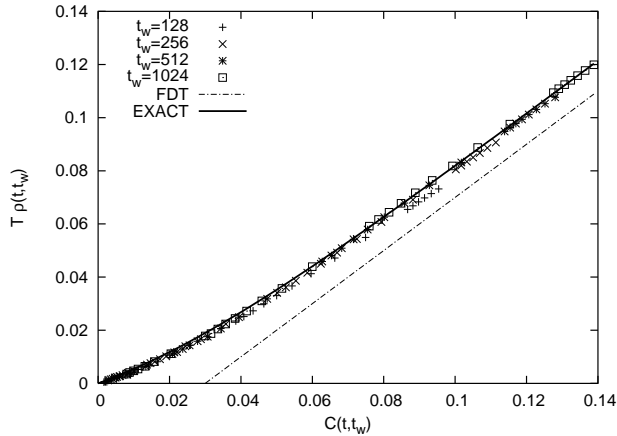}    
\caption{Parametric plot of the integrated response function $T
 \rho(t,t_w)$ as 
 a function of ${\cal C}(t,t_w)$ for different waiting times $t_w$ and
 $T=1.1\,T_g$.  The solid line is the
 result for the pure case as given by Eq. (\ref{X_pure}) and does not
 contain any fitting parameter. The dashed line shows the slope
 corresponding to the non-violated FDT.}\label{FDR_Pure}
\end{figure}

In order to characterize the deviation from the equilibrium, we
compute in this section the FDR $X(t,t_w)$
(\ref{Def_FDR}). For $T>T_g$ the disorder is irrelevant and 
the FDR is expected to be identical to the FDR of the
pure case, which we therefore consider first: in the pure 
model the aucorrelation and the response function can be computed
analytically. In the out of equilibrium regime $t_w < t \ll L^2$ 
(remembering that $z=2$ for the pure case) one has \cite{cugliandolo_pure}:
\begin{eqnarray}
&&{\cal R}_{\text{pure}}(t,t_w) = \frac{1}{T_g(2
  \pi)^2}\frac{1}{t-t_w} \quad, \quad 
  t>t_w \nonumber \\
&&{\cal C}_{\text{pure}}(t,t_w) = \frac{T}{T_g (2 \pi)^2}
  \ln{\left(\frac{t+t_w}{|t-t_w|} \right)} \label{Non_eq_pure} 
\end{eqnarray}
Using these expressions (\ref{Non_eq_pure}) together with
(\ref{Def_FDR}), one obtains that $X(t,t_w) \equiv X({\cal
C}(t,t_w))$ which allows to write the relation
defining the FDR (\ref{Def_FDR}) in an integrated form using the
definition of $\rho(t,t_w)$ (\ref{Def_rho})
\begin{eqnarray}
&&T \rho_{\text{pure}}(t,t_w) = \int_0^{t_w} ds X_{\text{pure}}({\cal
    C}_{\text{pure}}(t,s)) 
\partial_s {\cal C}_{\text{pure}}(t,s) \nonumber \\
&&= \hat{X}_{\text{pure}}({\cal C}(t,t_w)) -
\hat{X}_{\text{pure}}({\cal C}_{\text{pure}}(t,0))  \label{Def_Int_FDR} 
\end{eqnarray}
with $\partial_u \hat{X}_{\text{pure}}(u) = {X}_{\text{pure}}(u)$.
${\cal C}_{\text{pure}}(t,0)$ is expected to be small one can extract
${\hat X}_{\text{pure}}({\cal C}(t,t_w))$ from the slope of the curve
$T\rho_{\text{pure}}(t,t_w)$ versus ${\cal C}_{\text{pure}}(t,t_w)$ in
a parametric plot, provided $t_w$ is sufficiently large such that the
curves for different $t_w$ collapse.  In Fig. (\ref{FDR_Pure}) this
parametric plot $T\rho_{\text{pure}}$ versus ${\cal C}_{\text{pure}}$
is shown. For large values of ${\cal C}_{\text{pure}}$ one expects to
recover the FDT, and a slope of value unity. On the other hand, as
${\cal C}_{\text{pure}}$ decreases all these curves converge to a same
master curve $X_{\text{pure}}(C)$, which, using (\ref{Non_eq_pure})
can be exactly computed for the pure model
\begin{eqnarray}\label{X_pure}
\hat{X}_{\text{pure}}(C) = \gamma \ln{\frac{e^{\frac{C}{\gamma}} + 1 }{2}}
 \quad , \quad 
 \gamma = \frac{T}{(2 \pi)^2 T_g}
\end{eqnarray} 
As one can see in Fig. \ref{FDR_Pure}, our numerical results are in
good agreement with the exact calculation. An important point is that
the slope at the origin gives the asymptotic value of the FDR
$X^{\infty}_{\text{pure}}$, Eq. (\ref{xinf}) such that 
$T\rho_{\text{pure}}(t,t_w) \sim X^{\infty}_{\text{pure}}\cdot 
{\cal C}_{\text{pure}}(t,t_w)$ when ${\cal C}_{\text{pure}}(t,t_w) \to
0$. As is obvious from Eq. (\ref{X_pure}) for the pure model 
one has $X^{\infty}_{\text{pure}} = 1/2$, the random walk value 
\cite{cugliandolo_pure}, {\it independent} of the temperature.

For a finite size system, one expects to recover the equilibrium
dynamical regime for large but finite waiting times $t_w \gg t_{EQ}$
and in particular the restoration of the FDT (\ref{FDT}) reflected by $X(t,t_w)
=1$.  Therefore, as predicted by the analytical solution, the
parametric curve of $T\rho$ vs. ${\cal C}$ will progressively move to
the right with increasing $t_w$ converging in equilibrium
($t_w\to\infty$) to a straight line passing through the origin.

\begin{figure}
\includegraphics[width=\figwidth]
 {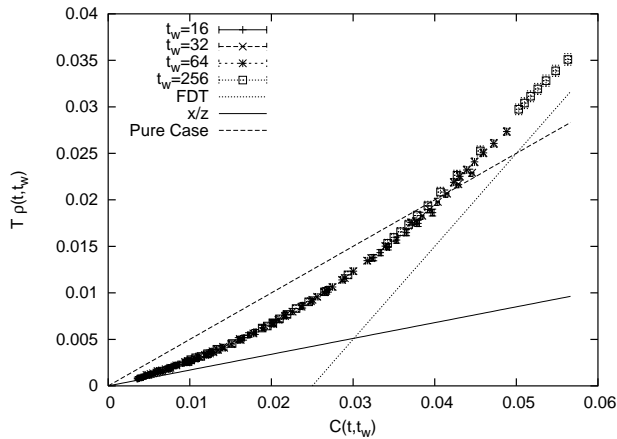}    
\caption{Parametric plot of the integrated response function $T
 \rho(t,t_w)$ as 
 a function of  
 ${\cal C}(t,t_w)$ for different waiting times $t_w$. Here $T=0.47
 \,T_g$. The solid line corresponds to a value of $X^{\infty} = 1/z$
 (\ref{Rel_X_z}) although the dashed one corresponds to $X_{\infty} =
 1/2$, thus showing a clear deviation from the pure case. The dotted
 line shows the slope corresponding to FDT.}\label{FDR_Dis_0.3}  
\end{figure}

\begin{figure}
\centering
\includegraphics[width=\figwidth]{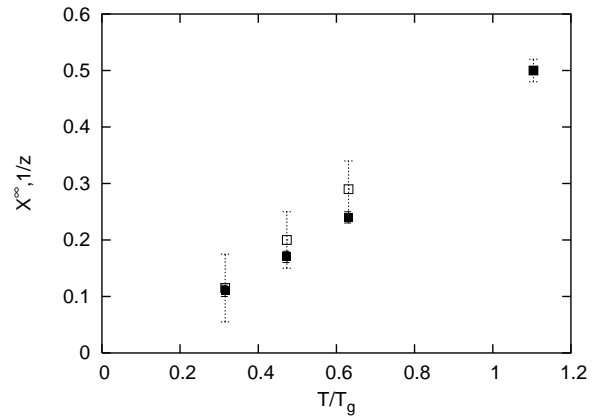}
\caption{Comparison between $X^{\infty}$, open symbols and $1/z$, filled
 symbols. The value of $1/z$ for $T=1.1\,T_g$ shown here is
 the exact one.} \label{Xinf_z} 
\end{figure}

We now turn to the case $T<T_g$ when the disorder is relevant. Given
the $t/t_w$ scaling forms we have obtained for ${\cal C}(t,t_w)$
(Fig. \ref{RescAutoCorrel0.4}) and for $\rho(t,t_w)$
(Fig. \ref{RescResp0.3}) one expects also in the disordered case to
have $\rho(t,t_w) \equiv \hat{X}(C(t,t_w))$. Indeed, as shown in
Fig. \ref{FDR_Dis_0.3} the parametric plot $T\rho$ vs. ${\cal C}$ for
different $t_w$ is qualitatively similar to the curve obtained for the
pure case. In particular, the property $\rho(t,t_w) \equiv
\hat{X}(C(t,t_w))$, together with Eq. (\ref{AutoCorrel_Coars}) yields,
in the nonequilibrium regime
\begin{eqnarray}
X(t,t_w) \equiv X \left( \frac{{\cal L}(t)}{{\cal L}(t_w)}\right)
\end{eqnarray}
Moreover, our data (Fig. \ref{FDR_Dis_0.3})
are consistent with a finite limiting value (as defined in Eq.(\ref{xinf}))
$X^{\infty} > 0$ also in presence of disorder
although the asymptotic value of this quantity is very difficult to
estimate numerically. This fact is qualitatively in agreement with RG
predictions. In contrast to the pure model, and according to
\cite{schehr_co_prl}, this value $X^{\infty}$ depends continuously on
$T$ as
\begin{eqnarray}
X^{\infty} = \frac{1}{z} + {\cal O}(\tau^2) \label{Rel_X_z}
\end{eqnarray}
close to $T_g$ \cite{schehr_co_prl}.  Although a precise comparison
with this RG predictions in the vicinity of $T_g$, where the
deviations from the pure case are expected to be small, is difficult
at this stage (requiring a study on longer time scales) one can see in
Fig. \ref{FDR_Dis_0.3} and Fig. \ref{Xinf_z}, that our data are still
in a reasonable agreement with the one loop relation (\ref{Rel_X_z}).

\section{COARSENING OR GROWING FLUCTUATIONS?}

The behavior we obtained for the $2$-point correlation function
$C(r,t)$ allowed to identify a growing length scale ${\cal L}(t)$ on
which the system gets equilibrated. To go further, one would like to
relate this length scale ${\cal L}(t)$ to the size of spatially
correlated structures like domains or droplets.  We first explored the
idea that at low temperature, the nonequilibrium dynamics could be
understood as a coarsening process reflected in a spatially growing
correlation with the ground state (GS).  Interestingly, computing the
GS of the SOS model on a disordered substrate (\ref{Def_SOS}) is a
minimum cost flow problem for which exits a polynomial algorithm and
can therefore be computed exactly \cite{rieger_gs_sos,
middleton_gs_sos}.  After determining one GS $n_i^0$ (note that the
GS, which is computed with free boundary conditions,
is infinitely degenerated since a global shift of all heights by an
arbitrary integer is again a GS), we define for each time $t$ the
height difference $m_i(t) = n_i(t) - n_i(0)$ and identify the
connected clusters (domains) of sites with identical $m_i(t)$ using a
depth-first search algorithm. Notice that for the comparison with
the Ground State, the Monte Carlo simulations are performed here 
using free boundary conditions. 

\begin{figure*}[ht]
\centering
\includegraphics[scale=1.]
 {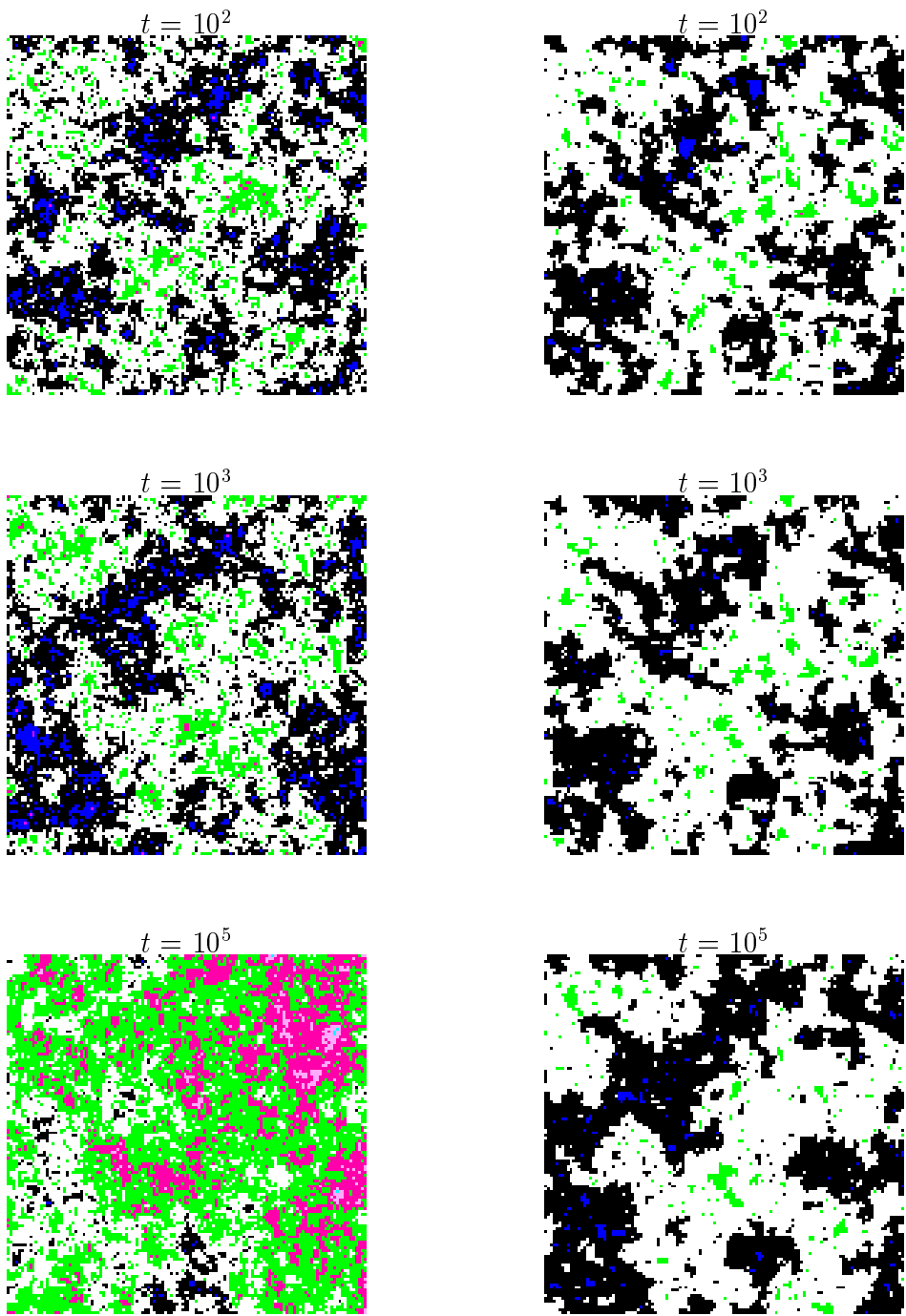}    
\caption{Snapshot of the height field relative to the ground state
 $m_i(t) = n_i(t) - n_i^0$ for $T > T_g$ on the left panel and $T=0.47\,
T_g$ on the right panel. Different colors correspond to different
 values of $m_i(t)$ : $m_i(t) = -2$ in green, $m_i(t) = -1$ in white,
 $m_i(t) = 0$ in black and $m_i(t)=+1$ in blue and so on.
 Note that for $T>T_g$ the configuration at $t=10^5$ is already
 decorrelated from the one at $t=10^3$, whereas for$T<T_g$ large
 domains in white and black persist and change only slowly in time.
}\label{CoarsPict} 
\end{figure*}

In Fig. \ref{CoarsPict} we show snapshots of these domains for
$T>T_g$ on the left panel and $T<T_g$ on the right one.

Starting from a random initial configuration one can for $T<T_g$ very
quickly ($t\lesssim100$) identify large domains that evolve only very
slowly at later times. On the other hand for $T>T_g$ the
configurations decorrelated very quickly in time. To make this analysis
more quantitative, we determined the cluster size distribution
$P_{\text{th}}(S,t)$ for one realization of the
disorder (and for different realizations of the thermal noise).

\begin{figure}
\centering
\includegraphics[width=\figwidth]{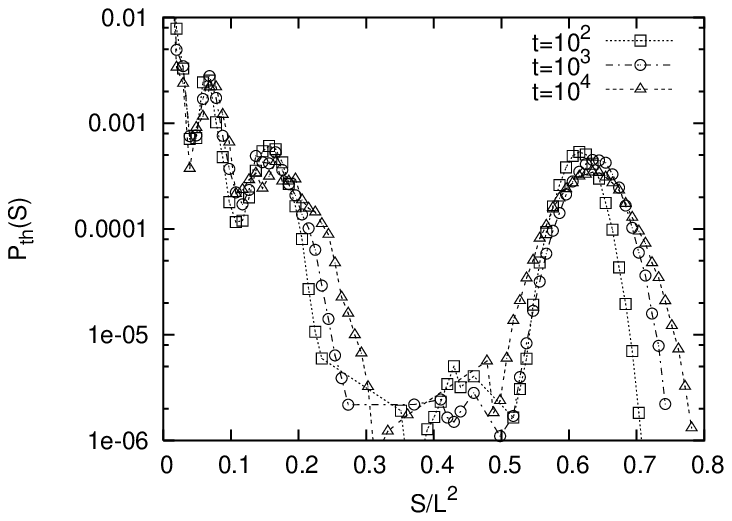}   
\caption{Size distribution $P_{\text{th}}(S,t)$ (see definition in the
 text) for different times $t$. Here $T = 0.47
 T_g$.}\label{Dist_Therm}
\end{figure}

As shown on Fig. \ref{Dist_Therm}, $P_{\text{th}}(S,t)$ starts to
develop a peak at a rather large value $S^*(t)$ on the earlier stage of
the dynamics (this peaks also develops if we start with a random
initial configuration). It turns out that $S^*(t)$ is the size of the
largest connected flat cluster of the ground state configuration
$n_i^0 = C^{\text{st}}$. On the time scales presented here, as time
$t$ is growing, this peak remains stable $S^*(t) \simeq
C^{\text{st}}$, implying that the system is {\it not} coarsening.
At later times, as suggested by simulations on
smaller systems, this peak progressively disappears and the
distribution becomes very flat. We also checked that the mean size of
these connected clusters is not directly related to ${\cal L}(t)$.

One has however to keep in mind that we are computing the {\it
connected} correlation functions, 
{\it i.e.} we measure the thermal fluctuations of the height profile
around its mean (typical) value $\langle h_i(t) \rangle$. Therefore, we
believe that these connected correlations are instead related to the
broadening of this ``stable'' peak (Fig. \ref{Dist_Therm}), {\it i.e.} the
fluctuations around this typical state at time $t$. The slow
evolution of the typical configuration, compared to the one of thermal
fluctuations around it, is corroborated by the one loop calculation
\cite{schehr_co_pre,schehr_co_prl} 
which shows that $\overline{\langle h_i(t) \rangle \langle h_i(t_w)
\rangle}$ decays as
\begin{eqnarray}
\overline{\langle h_i(t) \rangle \langle h_i(t_w)
\rangle} \sim \tau \left( \frac{t}{t_w}  \right)^{-1/2} + {\cal O}(\tau^2),
\end{eqnarray}
{\it i.e.} much slower than the connected one
(\ref{Def_Autocorrel}, \ref{one_loop_lambda}). 

\begin{figure}
\centering
\includegraphics[width=\figwidth]
 {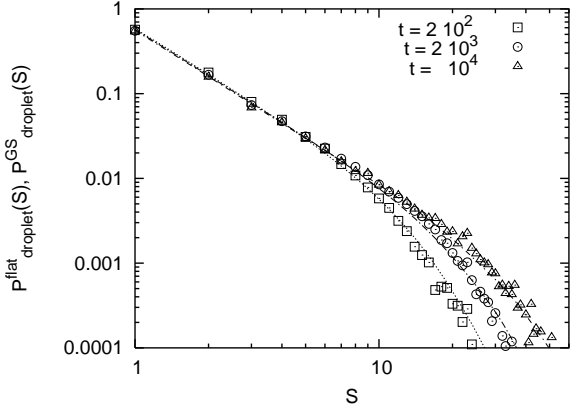}  
\caption{Size distribution $P^{\text{flat}}_{\text{droplet}}(S,t)$
 (symbols) for different times $t$. The solid lines represent 
$P^{\text{GS}}_{\text{droplet}}(S,t)$ (see the definition in the text)
 at the same corresponding 
 times. Here $T = 0.47\,T_g$.}\label{Dist_Act}
\end{figure}

To characterize more precisely the fluctuations of this
cluster, we have followed the following protocol: 
after a time $t_i \sim 100$ we store the configuration of the largest
connected cluster. Then, for each time $t$, we compute the distribution
$P^{\text{flat}}_{\text{droplet}}(S,t)$ of the
size of the connected clusters that were part of this cluster at
time $t_i$ but not at time $t$ (the subscript ``flat'' refers to the
{\it flat} initial condition). 
In Fig. \ref{Dist_Act}, we show a plot of
$P^{\text{flat}}_{\text{droplet}}(S,t)$ for a temperature $T=0.47
\,T_g$, for different times $t$.

It decays as a power law for small sizes $S$, and this power law
behavior extends to larger and larger values of $S$ as $t$ is growing. 
Although these data give already some interesting
insight on how the thermal fluctuations equilibrate in the system, 
it turns out to be very hard to obtain good statistics for
larger values of $S$ in this way. In order to perform a more precise
quantitative analysis of this distribution
we identify alternatively these ``droplets'' by initializing the
system in the ground state itself $n_i(t=0) = n_i^0$. At low temperature, 
and on the time scales explored here, one expects that 
the ground state represents a good
approximation of a typical configuration, {\it i.e.} $\langle n_i(t)
\rangle \simeq n_i^0$. Again we compute the distribution
$P^{\text{GS}}_{\text{droplet}}(S,t)$ of the
sizes of the connected clusters with a common value of $m_i(t)\ne0$.
As shown in Fig. \ref{Dist_Act}
$P^{\text{GS}}_{\text{droplet}}(S,t)$ determined 
in this way coincides very well with
$P^{\text{flat}}_{\text{droplet}}(S,t)$. Moreover, the calculation of
$P^{\text{GS}}_{\text{droplet}}(S,t)$ is much easier and allows for a more
precise analysis. 

\begin{figure}
\includegraphics[width=\figwidth]
 {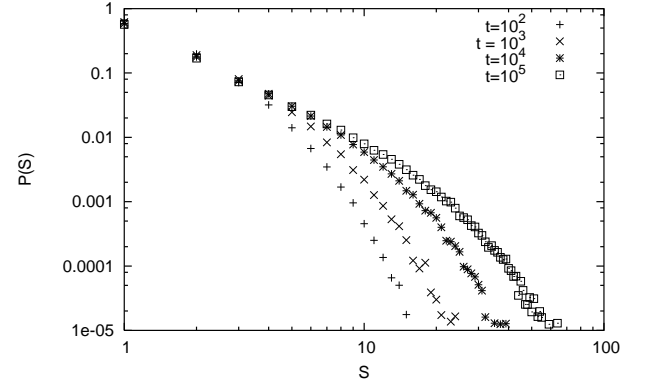}    
\caption{Distribution of the size of the clusters
  $P^{\text{GS}}_{\text{droplet}}(S,t)$ as a function of $S$ and for
  different times $t$. Here, the initial
 condition is the Ground State and $T=0.3 \,T_g$.}\label{Droplet_0.2} 
\end{figure}

\begin{figure}
\includegraphics[width=\figwidth]
 {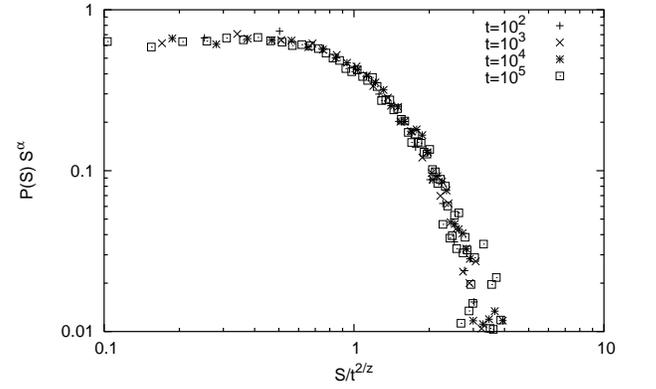}    
\caption{$S^{\alpha} P^{\text{GS}}_{\text{droplet}}(S,t)$ with $\alpha
 = 1.9 \pm 0.1$ as a function of $S/t^{2/z}$
 with $2/z = 0.26 \pm 0.03$. Here the initial condition is the Ground State
 and $T=0.3 \,T_g$.}\label{RescDroplet_0.2} 
\end{figure}

In Fig. \ref{Droplet_0.2}, we show a plot of
$P^{\text{GS}}_{\text{droplet}}(S,t)$ extending to larger times, for a
temperature $T=0.3 \,T_g$. It turns out, as shown in
Fig. \ref{RescDroplet_0.2}, that $P_{\text{droplet}}^{\text{GS}}(S,t)$
obeys the scaling form
\begin{equation}
P_{\text{droplet}}^{\text{GS}}(S,t) = \frac{1}{S^{\alpha}} {\cal
  F}_{\text{droplet}}^{\text{GS}} 
\left( \frac{S}{{\cal L}^2(t)}\right) \quad, \quad \alpha = 1.9 \pm 0.1 
\label{active_GS}  
\end{equation}
where $\alpha$ is independent of $T$ within the accuracy of our data
and ${\cal L}(t) \sim t^{1/z}$. The value of $z$ in (\ref{active_GS})
is in good agreement with the one extracted from the $2$-point
correlation function $C(r,t) = F(r/{\cal L}(t))$
(\ref{Scaling_Struct_Fact}). Furthermore, considering that each
``droplet'' of size $S > r^2$ gives a contribution to $C(r,t)$
proportional to $S$, one obtains, given the distribution
(\ref{active_GS}) with $\alpha = 2$
\begin{eqnarray}
&&C(r,t) \propto \int_{r^2}^{\infty} dS S
 P_{\text{droplet}}^{\text{GS}}(S,t) \\
&& \propto \ln{{\cal L}(t)/r} \quad, \quad {\cal L}(t)/r \ll 1 \nonumber 
\end{eqnarray}
which is consistent with the behavior we obtained in 
Fig. (\ref{RescStructFact0.3}) and Eq. (\ref{sos}).  

This scaling form 
(\ref{active_GS}) thus establishes a relation between ${\cal
  L}(t)$ and the typical size of compact excitation around a ``typical''
  configuration, evolving slowlier in time.

\section{CONCLUSION.}

In conclusion, we have performed a rather detailed analysis of the
nonequilibrium relaxational dynamics of the SOS model on a disordered
substrate (\ref{Def_SOS}), and of the related Cardy Ostlund model
(\ref{Def_SG}). Close to the superroughening temperature $T_g$ our results
for the autocorrelations, spatial correlations and response function
as well as for the fluctuation dissipation ratio (FDR) agree well with
the prediction of a recent one loop RG calculation
\cite{schehr_co_prl}, whereas deep in the glassy low temperature phase
substantial deviations occur. 

The aging features obtained perturbatively, characterized by a $t/t_w$
scaling of local correlation and response functions with a temperature
dependent decay exponent, carries over into the low temperature
regime, including a non-trivial temperature dependent fluctuation
dissipation ratio $X_{\infty}$ associated to these correlation and
response functions. The change in the low temperature behavior of
these quantities compared with the RG predictions turns out to be
contained in a change of the functional temperature dependence of the
dynamical exponent $z(T)$, which relates the age $t$ of the system
with a length scale ${\cal L}(t)$: $z(T)$ changes from a linear
$T$-dependence close to $T_g$ to a $1/T$-behavior far away from $T_g$.
This is, to our knowledge, the first clear indication of an activated
dynamics over logarithmic barriers in this marginal glass phase ({\it
i.e.} $\theta = 0$). Given the strong similarity of the behavior of
$z$ with the one found for the related model of a particle in a $1$-d
disordered potential with logarithmic correlations
\cite{marginal_freezing}, an open question remains whether this
dynamical crossover admits a static counterpart as found in that model
\cite{marginal_freezing}. 

The growing length scale ${\cal L}(t)$, increasing algebraically with
the age of the system, turned out to be connected to the typical size
of the fluctuations around metastable configurations with long life
time in which the system gets trapped immediately after a quench into
the low temperature phase. In contrast to a standard coarsening
process, where the growing length scale represents the typical size of
domains (which are identified as spatial regions strongly correlated
with one of the ground states of the systems) we encounter here a
scenario in which already soon after a temperature quench theses domains are
actually very large, but do not grow further and are destroyed by
fluctuations of increasing spatial extent. Moreover, these
fluctuations itself can be again identified as connected patches of
ground state, or droplets.

The emerging picture for the aging dynamics below the superroughening
transition within the glassy low temperature phase thus differs from
various well established aging sceanrios in glasses, spin glasses and
other disordered systems: As pointed out above the approach to
equilibrium is not a coarsening process as it occurs in other
disordered systems like the random ferromagnet \cite{random-ising}. It
also differs from the aging process encountered in finite dimensional
spin glasses which also display coarsening
\cite{kisker_noneq_sg,parisi-prl} with doamins that can
straighforwardly be identified due to the existence of the
Edwards-Anderson order parameter. On the other hand the aging scenario
revealed for this system appears to be far from being as complex as in
mean-field spin glasses \cite{leto_leshouches}: It is more reminiscent
of the dynamics of a random walk in a one-dimensional energy
landscape, the Sinai-model, in which the walker displacement also
increases only logarithmically with time due to the existence of deep
traps with exponentially long trapping times \cite{sinai}. 

With regard to our observation that these traps in the disordered SOS
model can be identified with configurations roughly made of large
patches of the ground state it is tempting to describe the aging
process here as a diffusion in a coarse grained configuration space
consisting of height profiles composed like a jigsaw puzzle of ground
state domains of optimzed shape (most probably flat pieces of constant
height with energy minimizing boundaries). The escape from a deep
energy minima proceed, according to what our numerical analysis
indicates, via the thermal activation of larger and larger patches,
each intermediate configuration again being metastable with some
finite survival time. This process is reminiscent of the
energy-well-within-energy-well picture proposed in \cite{balents-etal}
and in our view further studies would be worthwhile to develop this
analogy in more detail.

\acknowledgments
GS aknowledges D. Dominguez and A. Kolton for useful discussions at
the earliest stage of this work and acknowledges the financial support
provided through the European Community´s Human Potential Program
under contract HPRN-CT-2002-00307, DYGLAGEMEM.

\appendix

\section{Comparison with RG calculations near $T_g$}\label{Comp_RG}

In this appendix we establish the connexion between the quantities (in
Fourier space) computed analytically in \cite{schehr_co_prl} and the
ones in real 
space computed numerically in the present paper. We give here the
details for the connected autocorrelation function ${\cal C}(t,t_w)$
(\ref{Def_Autocorrel}), the extension to the integrated response
$\rho(t,t_w)$ (\ref{Def_Rho}) being then straighforward. 
In \cite{schehr_co_prl}, the analytical predictions focused on
the following connected correlation function
\begin{eqnarray}\label{Correl_Fourier}
{\cal \hat{C}}^q(t,t_w) = \overline{\langle \hat h_q (t) \hat
  h_{-q}(t_w)\rangle - \langle 
\hat h_{q}(t)\rangle \langle \hat h_{-q}(t_w) \rangle}
\end{eqnarray} 
where $\hat h_q(t)$ is the Fourier transform, w.r.t. space variable,
of the field $h_i(t)$ (\ref{Def_SOS}). Using RG along the line of fixed
points near $T_g$, this correlation function (\ref{Correl_Fourier})
was computed up to order ${\cal O}(\tau^2)$. It takes the following
form 
\begin{eqnarray}\label{theta_RG}
&&{\cal \hat{C}}^q(t,t_w) = \frac{T}{q^2}
\left(\frac{t}{t_w}\right)^{\theta_C}  F_C(q^z(t-t_w),t/t_w) \\
&& \theta_C = e^{\gamma_E} \tau + {\cal O}(\tau^2)\nonumber
\end{eqnarray}  
where $\gamma_E$ is the Euler constant, given in the text, and with
the asymptotic behavior in the large time separation limit
\begin{eqnarray}\label{Asympt_Correl_Fourier}
F_C(v,u) = \frac{F_{C \infty}(v)}{u} + {\cal O}(u^{-2})  
\end{eqnarray}
The connected autocorrelation function ${\cal C}(t,t_w)$
(\ref{Def_Autocorrel}) we compute here is related to
(\ref{Correl_Fourier}) through 
\begin{eqnarray}\label{Expl_Rel}
&&{\cal C}(t,t_w) = \int \frac{d^2 q}{(2 \pi)^2} {\cal
    \hat{C}}^q(t,t_w) \\
&&= \frac{T}{(2 \pi)^2} \left(\frac{t}{t_w} \right)^{\theta_C} \int
\frac{d^2 q}{q^2} F_C(q^z(t-t_w),t/t_w) \nonumber
\end{eqnarray}
Perfoming the change of variable $v = q^z(t-t_w)$, (\ref{Expl_Rel})
becomes
\begin{eqnarray}
{\cal C}(t,t_w) = \frac{T}{2 \pi z} \left(\frac{t}{t_w}
\right)^{\theta_C} \int_0^{\infty} \frac{dv}{v} F_C(v,t/t_w)
\end{eqnarray}
where we have taken the IR (resp. the UV) cutoff to $0$ (resp. to $\infty$)
and checked the convergence of the integral over $v$. Using the
asymptotic behavior (\ref{Asympt_Correl_Fourier}) one obtains (the
remaining integral over $v$ being well defined) in the
large time separation limit $t \gg t_w$ 
\begin{eqnarray}\label{Asympt_Autocorrel_RG}
{\cal C}(t,t_w) \sim  \frac{T}{2 \pi z} \left(\frac{t}{t_w}
\right)^{\theta_C-1}\int_0^{\infty} \frac{dv}{v} F_{C\infty}(v)  
\end{eqnarray}
which, given the value of $\theta_C$ (\ref{theta_RG}), 
leads to the following one loop result for the decay exponent
$\lambda/z$ (\ref{AutoCorrelAsympt}):
\begin{eqnarray}
\lambda/z = 1 - e^{\gamma_E} \tau + {\cal O}(\tau^2)
\end{eqnarray}
given in the text in Eq.(\ref{one_loop_lambda}).

\end{document}